\documentclass[review]{elsarticle}

\usepackage{lineno,hyperref}
\modulolinenumbers[5]

\usepackage{siunitx}

\makeatletter
\def\ps@pprintTitle{%
	\let\@oddhead\@empty
	\let\@evenhead\@empty
	\def\@oddfoot{\reset@font\hfil\thepage\hfil}
	\let\@evenfoot\@oddfoot
}
\makeatother

\begin{document}

\begin{frontmatter}

\title{Geomagnetic Storm Main Phase Effect on the Equatorial Ionosphere over Ile--Ife as measured using GPS Observations}

\author[mymainaddress]{Ayomide O. Olabode\corref{mycorrespondingauthor}}
\cortext[mycorrespondingauthor]{Corresponding author}
\ead{aolabode@oauife.edu.ng}

\author[mymainaddress]{Emmanuel A. Ariyibi}

\address[mymainaddress]{Department of Physics and Engineering Physics, Obafemi Awolowo University, Ile--Ife, Nigeria}

\begin{abstract}
The effect of the main phase of two intense geomagnetic storm events which occurred on August 5 -- 6 and September 26 -- 27, 2011 on the equatorial ionosphere have been investigated using Global Positioning System (GPS) data obtained from an Ile-Ife station (geomagnetic lat. \ang{9.84}N, long. \ang{77.25}E, Dip \ang{7.25}N). Total Electron Content (TEC) profiles during the main phase of the two geomagnetically disturbed days were compared with quiet time average profiles to examine the response of the equatorial ionosphere. The results showed that the intensity of both storm events during the main phase which occurred at night-time correlated well with a strong southward direction of the z-component of the Interplanetary Magnetic Field (IMF-Bz) and Solar Wind Speed (Vsw), with the Disturbance storm time (Dst) profile showing multiple step development. TEC depletion was observed during the main phase of the August 5 -- 6, 2011 storm event with TEC recording a maximum value of 9.31 TECU. A maximum TEC value of 55.8 TECU was recorded during the main phase of the September 26 -- 27, 2011 storm event depicting TEC enhancement. Significant scintillation index value of 0.57 was observed when the main phase started on August 5 -- 6, 2011 followed by a prolonged suppression while there was less significant scintillation impact on September 26 -- 27, 2011 with a maximum value of 0.33. The present study show that rapid energy input from solar wind during geomagnetic storm events effect large variations in TEC and significant scintillation phenomenon in the equatorial ionosphere over Ile-Ife, Nigeria.
\end{abstract}


\end{frontmatter}


\section{Introduction}
\label{sec1}
Efficient transfer of energy from solar wind outbursts from active regions of the Sun into the earth’s magnetosphere generates major disturbances referred to as geomagnetic storms \cite{Dujanga2013}. Such disturbances translate into perturbations of the charged particles of the ionosphere and can cause detrimental effects on radio-wave signals and satellite communications. These effects have been reported to be more profound at equatorial regions of the earth's ionosphere where large variations in ionospheric parameters have been observed \citep{Ariyibi2013,Fayose2012,Jain2010}, mainly because of the existence of a ring current flowing in the equatorial plane.

Typically, a geomagnetic storm is characterized by three phases: an initial phase, a main phase and a recovery phase \cite{DeJesus2012}. The main phase is the defining feature of a geomagnetic storm and can be described by a significant depression of the magnetic field on the Earth’s surface. This is manifested as a significantly southward-driven (z-component of) interplanetary magnetic field (IMF-Bz), thus intensifying energy input in the upper atmosphere \cite{Kamide2001,DeJesus2012,Dujanga2013} and causing the ring current to become energized.

A geomagnetic storm is usually identified by the strength of the low latitude magnetic disturbance storm time index (Dst) which is the measure of the intensity of the magnetospheric ring current. The degree of the Dst main phase decrease is a measure of the intensity of the storm and may be characterized as intense/super, moderate and weak for the Dst $ < –100 $ nT, –100 nT $ < $ Dst $ < –50 $ nT, –50 nT $ < $ Dst $ < –30 $ nT respectively \cite{Abdu2012}. The Planetary K (Kp) index which is derived from a K index, can also be used to describe global geomagnetic activity. K index is a quasi-logarithmic index characterizing the 3-hourly range intransient magnetic activity relative to the regular \lq\lq quiet-day" activity for a single site location \cite{Achem2013}. When Kp $ < $ 5, it describes a non-occurrence of geomagnetic activity. Kp = 5, Kp = 6, Kp = 7, Kp = 8 and Kp = 9 respectively indicate a minor, moderate, strong, severe and extreme geomagnetic activity.

To describe the effects of geomagnetic storms on the ionosphere, GPS observations can provide us measurements from which ionospheric parameters TEC and $ S_4 $ can be estimated. TEC is defined as the number of electrons in a column of 1 $ m^2 $ cross-section, from the height of the GPS satellite to the receiver on the ground \cite{Dujanga2013}. It provides an overall description of the ionosphere and can be expressed as:
\begin{equation}
TEC = \int_{Rx}^{Tx} Ne\ ds
\end{equation}
where TEC is measured in TEC units (TECU) with 1 TECU = $ 10^{16} $ electrons/m$ ^2 $. $T_x $ and $ R_x $ represent the GPS satellite and ground receiver positions respectively in km. GPS satellites are observed along oblique signal paths which pierce the ionosphere at ionospheric pierce points (IPPs). The height of the IPP corresponds to the height typically associated with the peak electron density in the ionosphere, $ \approx 350 $ km for this study. 

Ionospheric scintillation is characterized by rapid fluctuations in the amplitude and phase of transionospheric radio signals due to variations in the local index of refraction along the propagation path. The fluctuations in the signal amplitude are quantified by $ S_4 $ (rapid fluctuations in signal to noise ratio (C/No) are interpreted as modulation of the signal due to ionospheric scintillation, e.g. $ S_4 $ = 0.0 indicates no modulation whereas $ S_4 $ = 1.0 indicates 100\% modulation) \cite{Carrano2008}.The $ S_4 $ is the statistical measure of the intensity of the amplitude scintillation of the L1 frequency (1575 MHz) GPS signals. It is the ratio of the signal intensity standard deviation and the signal intensity mean and can be expressed as:
\begin{equation}
S_4 = \sqrt{\frac{(<I^2> - <I>^2)}{<I>}}
\end{equation}

In this paper, ionospheric observations obtained from GPS measurements during the intense geomagnetic storm events that occurred on August 6, 2011 and September 26, 2011 are presented to discuss the variability of the equatorial ionosphere over Ile-Ife, Nigeria during a storm.

\section{Measurement Station Set-Up and Data}
\label{sec2}
The geomagnetic storms that occurred on August 5 -- 6 and September 26 -- 27, 2011 for which data were available at the Ile-Ife Station was considered for investigation. Solar interplanetary and geomagnetic parameters: The IMF-Bz, Solar wind velocity (Vsw) and Kp index for the period under investigation were obtained from the 1-h averaged Operating Missions as Nodes on the Internet (OMNI) database available at the online NASA Space Data Facility. The level of geomagnetic activity during the periods examined is indicated by the Dst index and was obtained from the World Data Centre for Geomagnetism, Kyoto.

GPS measurements have been obtained using the high data-rate NovAtel GSV4004B GPS-SCINDA receiver installed at the Department of Physics and Engineering Physics, Obafemi Awolowo University, Ile-Ife, Nigeria (geom. lat. \ang{9.84}N, long. \ang{77.25}E, Dip \ang{7.50}N). The GPS-SCINDA is a real-time GPS data acquisition and ionospheric analysis system which computes ionospheric parameters TEC and $ S_4 $ using the full temporal resolution of the receiver. The TEC is computed from the combined L1 (1575 MHz) and L2 (1228 MHz) pseudoranges and carrier phase and it involves the estimation of the absolute TEC by first levelling the phases to the pseudorange to give the relative TEC and further estimating/removing instrumental biases \cite{Carrano2006}. 

The TEC and $ S_4 $ have been estimated from ionospheric statistics file (*.scn file) provided by GPS-SCINDA using the GPS-TEC Analysis Software Version 2.9. This program performs a set of algorithm on input GPS-SCINDA observation data, and provides vertical TEC and $ S_4 $ plots as well as ASCII output files which are then obtained for further data presentation, analysis and discussion. To eliminate multipath and higher horizontal gradient effects, the program reads navigation file for the observation date to calculate elevation angles of the satellites; an elevation angle limit of \ang{20} was applied. 

To specifically examine variations in ionospheric TEC signatures during the main phase of the storm events, TEC values for ten (10) days of \lq quiet' geomagnetic activity were obtained for each month of August and September 2011; (observed) storm-time TEC deviation from monthly average quiet TEC was then estimated using the equation \ref{eq3} below:
\begin{equation}
\Delta TEC = \frac{TEC_{obs} - TEC_{qav}}{TEC_{qav}} \times 100
\label{eq3}
\end{equation} 
where $ TEC_{obs} $ is the observed disturbed value of the TEC and $ TEC_{obs} $ is the average quiet TEC value. All TEC values in this study are presented in TEC units.

\section{Results and Discussion}
\label{sec3}
\subsection{Solar Interplanetary and Geomagnetic Observations}
\label{subsec1}
\subsubsection{Observations for August 5 -- 6, 2011 Geomagnetic Storm}
\label{subsubsec1}
\begin{figure}[hbtp]
	\centering
	\includegraphics[width=0.7\linewidth]{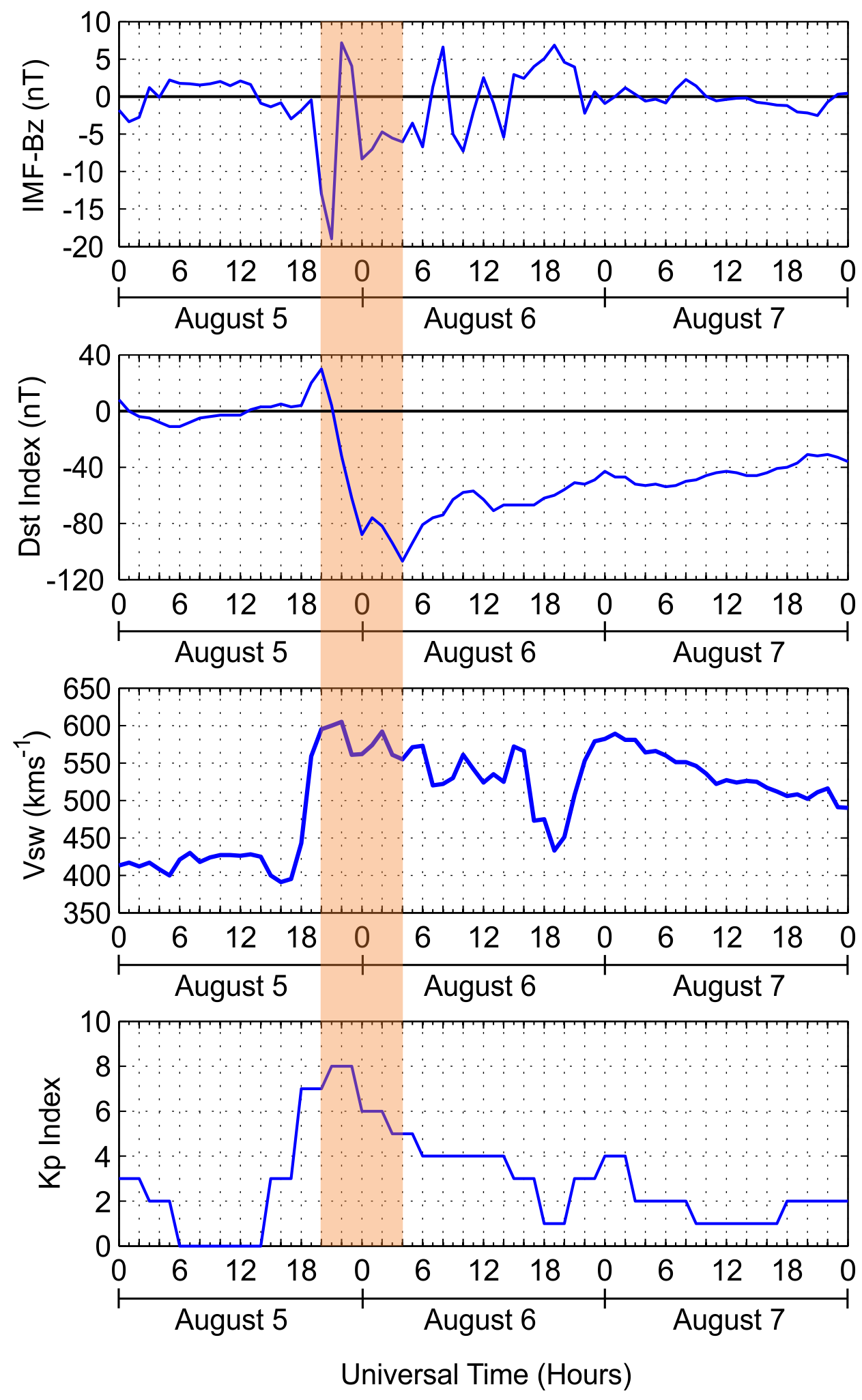}
	\caption{Interplanetary and Geomagnetic Observations for August 5--7, 2011.}
	\label{fig:1}
\end{figure}

On August 4, a significant M9 class solar flare which was produced at about 0400 UT propelled a Coronal Mass Ejection (CME) towards earth. The interplanetary event, corresponding to the August 6, 2011 geomagnetic storm is shown in Figure \ref{fig:1} for the period of August 5--7, 2011. The panels from top to bottom are: IMF-Bz, Dst Index, Vsw and Kp Index. 

As shown in Figure \ref{fig:1}, IMF-Bz recorded negative values for the first four hours of August 5 with minimum value of –3.37 nT at 0100 UT. After turning northwards at 0400 UT, it maintained positive values until 1300 UT when it began a gradual southward turning from 1400 UT with a value of –0.89 nT. IMF-Bz took a sharp sudden decrease at 1900 UT recording a value of –12.96 nT at 2000 UT, a corresponding sharp decrease was also observed in the Dst curve, at this point, the main phase of the geomagnetic storm event began.

The main phase of the storm event commenced at 2000 UT on August 5 with the sharp and rapid Dst index value drop from 30 nT to a minimum value of –88 nT by midnight. It recovered briefly to a value of –76 nT and persisted for another two hours before reaching a minimum value of –107 nT at 0400 UT on August 6. Dst Index values recovered steadily until 1100 UT (a value of –57 nT) but thereafter decreased to –71 nT at 1300 UT. Afterwards, the storm commenced a steady recovery phase. The main phase of this storm event occurred through midnight hours and endured over an eight-hour period.

The Dst profile shows that the main phase of this intense August 6 geomagnetic storm consists of a moderate storm (minimum Dst = –88 nT at 0000 UT) and a major storm (minimum Dst = –107 nT at 0400 UT on August 6) depicting a two-step storm development, that is, after the first minimum Dst is reached, a brief partial recovery intervenes to be followed by another minimum Dst (Kamide et al., 1998). Correspondingly, the presence of a shock in the interplanetary medium at 2000 UT is confirmed by the increase in Vsw from 559 $ kms^{-1} $ at 19:00 UT to $ \approx $ 600 $ kms^{-1} $ at 2000 UT on August 5. Vsw reached a maximum value of 605 $ kms^{-1} $ at 2200 UT on August 5. A minimum value of 555 $ kms^{-1} $ for solar wind speed was recorded at the end of the main phase of the storm event at 0400 UT when the Dst reached its maximum depression value. 
The storm registered 8 on the (0 -- 9) Kp Index scale of geomagnetic disturbance classifying it as an intense geomagnetic storm event. Its maximum value of 8 was recorded at 2100 UT on August 5 during the main phase of the storm event. It continued a slow decay reaching a minimum value of 5 at 0300 UT through 0400 UT on August 6 as the main phase of the storm event ended.

\subsubsection{Observations for September 26 -- 27, 2011 Geomagnetic Storm}
\label{subsubsec2}
\begin{figure}[hbtp]
	\centering
	\includegraphics[width=0.7\linewidth]{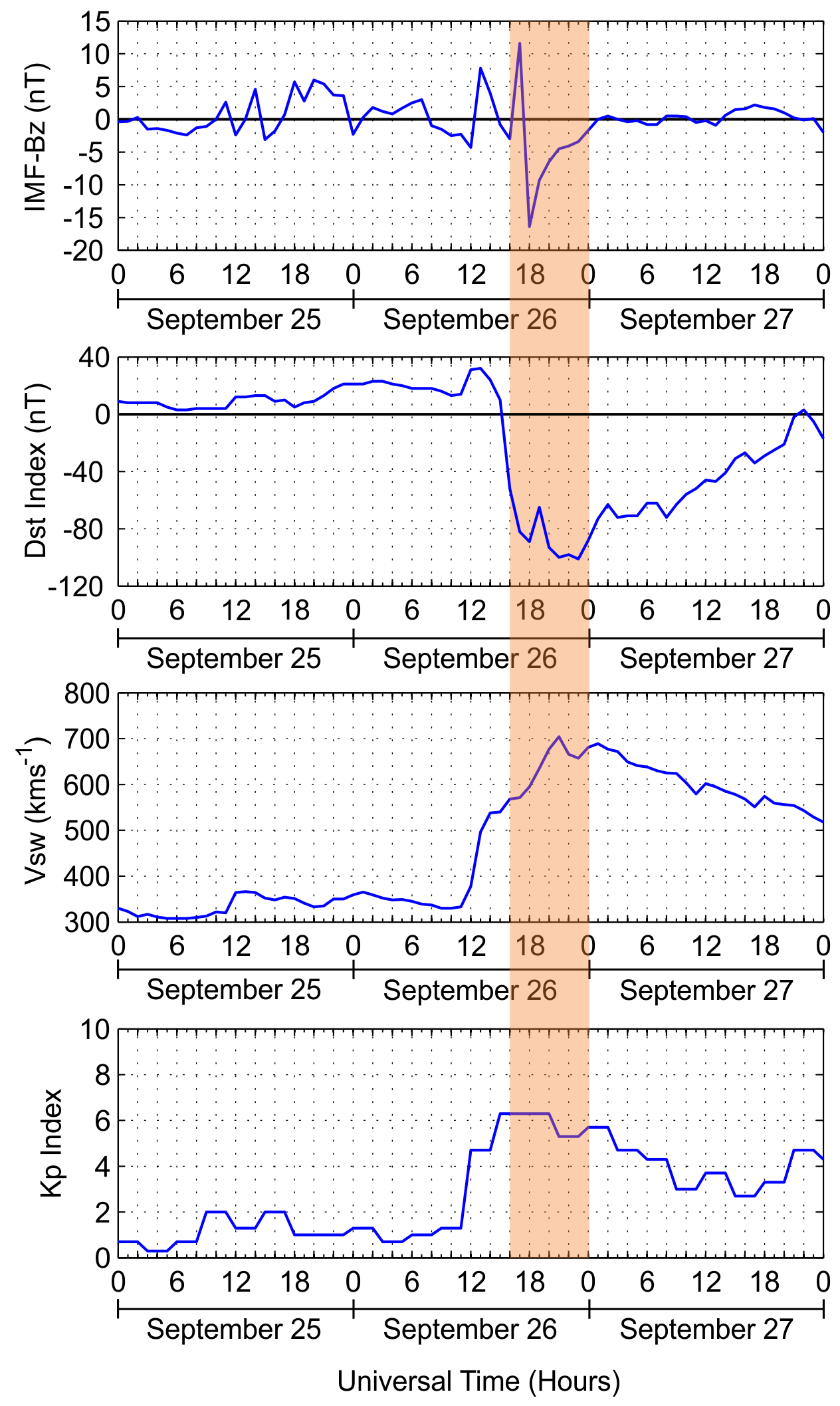}
	\caption{Interplanetary and Geomagnetic Observations for September 25--27, 2011.}
	\label{fig:2}
\end{figure}

To describe this storm event, IMF-Bz, Vsw and Kp Index data for the period of September 25--27, have been plotted in panels in Figure \ref{fig:2} in the same manner as in the previous section. 
On September 25, z-component of the IMF-Bz recorded small negative values between 0000 UT and 1600 UT. However, during this period, brief positive values were attained between 1100 UT and 1400 UT. Thereafter, IMF-Bz turned northward at 1700 UT and maintained positive values until 0700 UT on September 26. IMF-Bz turned southward afterwards with a value of –1.0 nT and continued a gradual decrease for the rest of the day with brief northward turnings observed during this period. A minimum value of –16.4 nT was recorded at 1800 UT on September 26, at this time, the Dst curve dropped to its first minima of –89 nT. A second minima of –100 nT was recorded at 2100 UT on September 26, a partial recovery intervenes and thereafter decreased very strongly to a third minima of –101 nT at 2300 UT. The main phase of this geomagnetic storm event occurred over an eight-hour period be-tween 1600 UT on September 26 and 0000 UT on September 27, 2011.

\cite{Kamide1998} stated that majority of intense storms undergo two-step development during the main phase as was also observed in the August 6 storm event. However, the Dst profile of the September 26 geomagnetic storm event does not appear to follow a two-step storm type. As shown in Figure \ref{fig:2}, three minima were recorded (two moderate ones and one major storm) depicting a triple-step development instead, this which is similar to what was reported by \cite{Chukwuma2010}. 

The interplanetary shock is identified by an abrupt increase in the Vsw from about 500 $ kms^{-1} $ to about 570 $ kms^{-1} $ at 1600 UT on September 26 when the main phase of the geomagnetic storm commenced. By 2100 UT on September 26, it reached a maximum value of 704 $ kms^{-1} $. While the Vsw recorded its peak value three hours after IMF--Bz recorded its minimum value of –16.4 nT at 1800 UT on September 26, Kp Index recorded its maximum value of 6, when IMF--Bz recorded its minimum value. Kp Index recorded this maximum value an hour before the main phase of the geomagnetic storm event commenced and remained constant for six hours. During the main phase of the geomagnetic storm event, it recorded its minimum value at midnight.

\subsection{Ionospheric Conditions during the Geomagnetic Storms}
\label{subsec2}
\subsubsection{Ionospheric Observations on August 5 -- 6, 2011 }
\label{subsubsec3}
\begin{figure}[hbtp]
	\centering
	\includegraphics[width=0.7\linewidth]{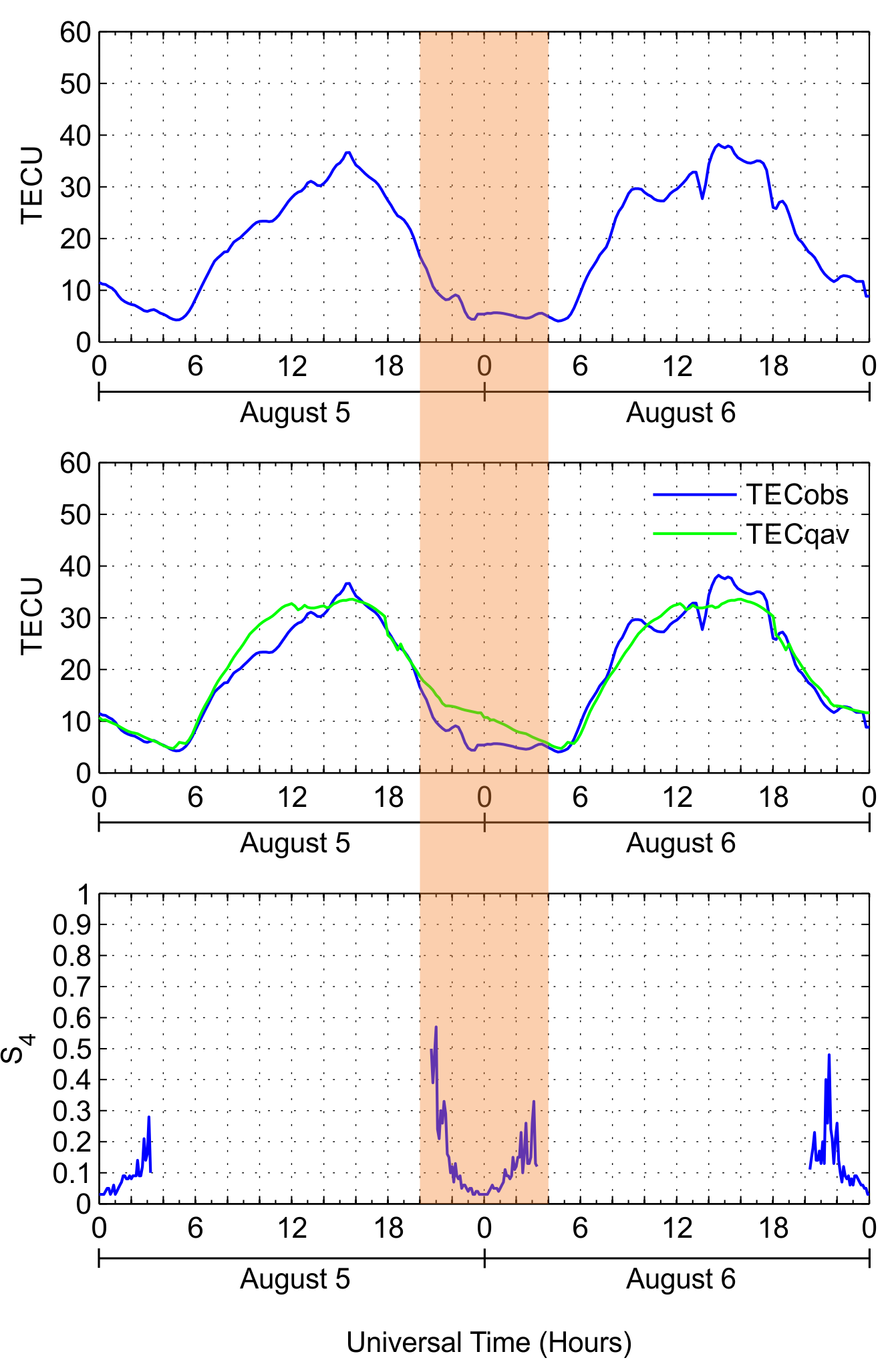}
	\caption{Observed Mean Ionospheric TEC and S4 for August 5 – 6, 2011}
	\label{fig:3}
\end{figure}

Figure \ref{fig:3} presents a plot of the TEC observed before, during and after the main phase of the geomagnetic storm event of Au-gust 5 -- 6, 2011 for the Ile-Ife Station. From the panels of plotted TEC, we observe a typical three segmented profile --- a build-up region, a daytime plateau and a decay region. On August 5, 2011 before sunrise, the build-up region attains a minimum value of about 4 TECU at 0500 UT after which it steadily increased to a daytime peak value of 35.6 TECU at 1500 UT. It experiences fluctuations between 1200 UT and 1500 UT with a slight depression noticeable at 1300 UT. It thereafter declined to a minimum value of 4.4 TECU at 2300 UT with a noticeable TEC depletion.
On August 6, 2011, TEC decreased in value from about 6 TECU at midnight to about 5 TECU at 0100 UT. It remained at this value for three hours after which it quickly increased to a value of 21.8 TECU at 0800 UT, it continued to increase steadily with multiple peaks until it attained a maximum daytime value of 37.5 TECU at about 1600 UT after which it commenced a steady decrease with noticeable wave-like fluctuations. A minimum value of 8.8 TECU was recorded at midnight. During the main phase of the geomagnetic storm which occurred between 2000 UT on August 5, 2011 and 0400 UT on August 6, 2011, TEC recorded values between 4 and 8 TECU. Compared with TEC values at the other two phases of the geomagnetic storm, TEC values during the main phase fell by about 80\%. 

To describe the effect of the August 5 -- 6, 2011 storm event on TEC at the Ile-Ife station, observed TEC (denoted as $ TEC_{obs} $) and average TEC for the 10 geomagnetically quietest days in August (denoted as $TEC_{qav} $) were plotted together as shown in the middle panel of Figure \ref{fig:3}. The storm effect on TEC was significant such that large deviations were noticeable. Except at 1600 UT when $TEC_{obs} $ profile peaked with a value of 37 TECU, the $TEC_{qav} $ recorded significantly larger values of TEC than $TEC_{obs} $. On August 5, the $TEC_{qav} $ daytime segment (0700 UT – 1400 UT) showed TEC values significantly larger than the daytime segment of the $TEC_{obs} $ profile. During this period, the $TEC_{qav} $ peaked at 1200 UT with a value of 32.7 TECU while $TEC_{obs} $ profile recorded a TEC value of 27.9 TECU. During the main phase of the geomagnetic storm (2000 UT on August 5 to 0400 UT on August 6), TEC depletions were noticeable. When the storm main phase commenced at 2000 UT, $TEC_{obs} $ recorded a value of 17 TECU while $TEC_{qav} $ recorded a value of 19 TECU. The variation became more pronounced around 2300 UT when the maximum variation was observed; while $TEC_{obs} $ recorded a value of 4.5 TECU, $TEC_{qav} $ recorded a value of 11.9 TECU. By the time the main phase ended at 0400 UT, the variation was still visible however, smaller --- $TEC_{obs} $ recorded a value of 4.4 TECU while $TEC_{qav} $ recorded a value of 5.1 TECU.

The bottom panel of Figure \ref{fig:3} shows the variation of the scintillation index. Satellite with Pseudo Random Number (PRN) 23 showed the most remarkable scintillation records during the main phase hours of the August 5 -- 6, 2011 geomagnetic storm event and was considered for interpretation. For the storm event that occurred on August 5 -- 6, 2011, $ S_4 $ values ranged between a minimum and maximum of 0.0 and about 0.57 respectively. Scintillation values greater than or equal to 0.3 are considered to imply a significant level of scintillation (Dashora and Pandey, 2005). The pronounced scintillation phenomenon is observed to have commenced with a value of 0.5 shortly before 2100 UT, by 2100 UT when the main phase of the geomagnetic storm event commenced, this value had risen to 0.57 corresponding to rapid fluctuations in TEC. This is followed by a remarkable suppression between 2200 UT on August 5 and 0200 UT on August 6 before it rose to a value of 0.33 around 0300 UT. This observation is most likely associated with the effect due to prompt penetration of an eastward electric field into the equatorial ionosphere and this is similar to the observation reported by Priyadarshi and Singh (2011).

\subsubsection{Ionospheric Observations on September 26 -- 27, 2011}
\label{subsubsec4}

\begin{figure}[hbtp]
	\centering
	\includegraphics[width=0.7\linewidth]{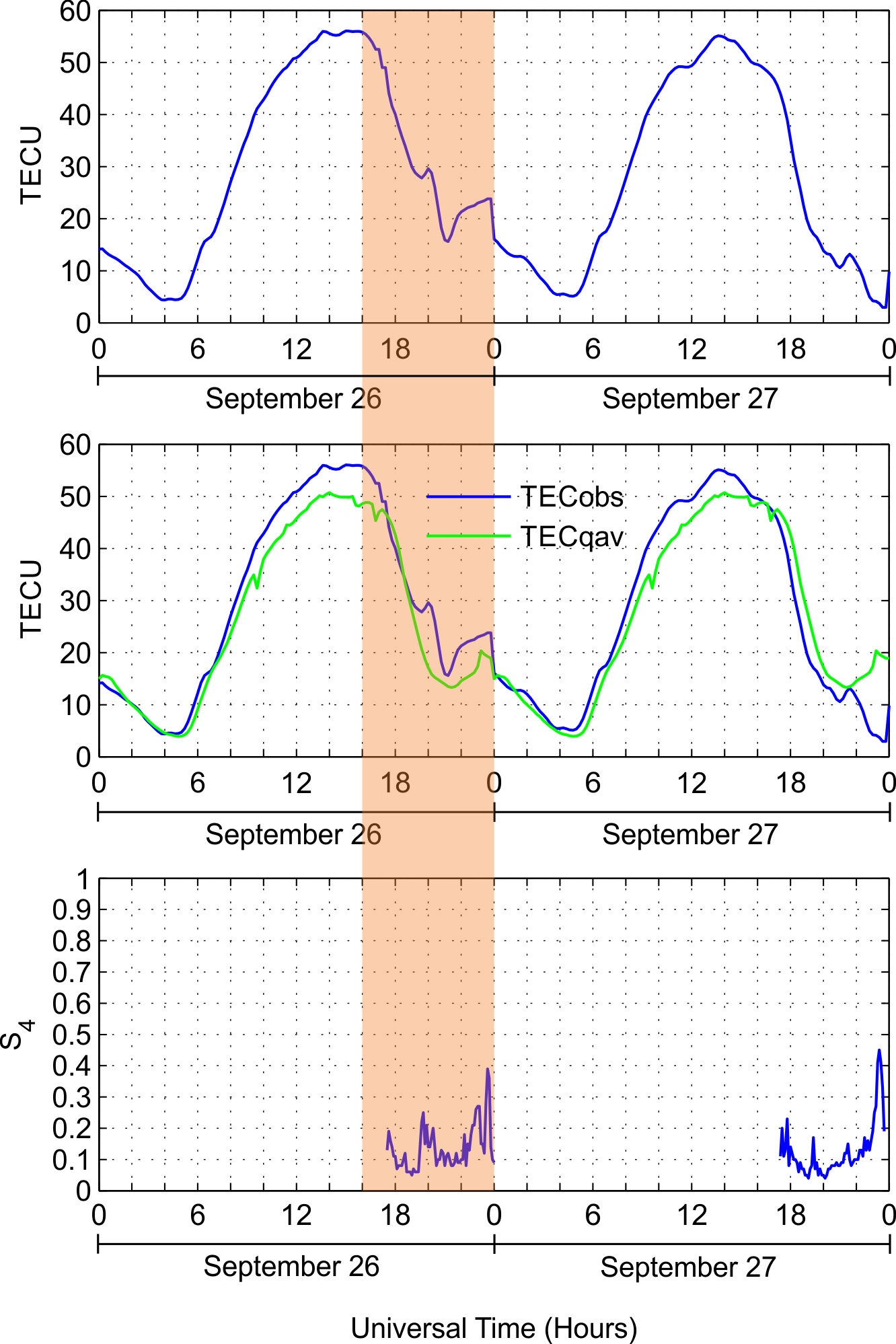}
	\caption{Observed Mean Ionospheric TEC and S4 for September 26 – 27, 2011}
	\label{fig:4}
\end{figure}

The observed mean TEC, the TEC variation from quiet TEC averages and $ S_4 $ index for the geomagnetic storm event of September 26 -- 27, 2011 for the Ile-Ife Station is presented from top to bottom respectively in the Figure \ref{fig:4}. Satellite with PRN 23 showed the most remarkable scintillation records and was considered for interpretation. A sharp fall in TEC value is observed at pre-sunrise hours on September 26, 2011 reaching a minimum value of about 3 TECU at around 0500 UT. A sharp morning rise is thereafter recorded with TEC reaching a daytime maximum value of $ \approx 60 $ TECU at 1400 UT forming a relatively stable daytime plateau region for about two hours following which the decay region commenced until midnight. An almost similar trend is observed on September 27. A remarkable scintillation effect was observed during the main phase which occurred between 1600 UT and midnight.
Noticeably, unlike the main phase of the August 6 storm event which occurred between the post-sunset hours of August 5 and pre-dawn hours of August 6, the main phase for September 26 (an Equinox month) occurred only during the post-sunset hours. Also, the morning rise and afternoon decay in TEC is sharper during the September 26 storm when compared to that of August 6 storm, this is similar to findings by \cite{Bhuyan2007}. Furthermore, the daytime peak of TEC values was observed to be higher during the September 26 storm. The TEC profile during this storm event also exhibited a more pronounced daytime plateau region than that of the August 6 storm event. 

TEC variation was observed to be significantly large during the day-time. On September 26, the average quiet-time TEC profile showed that daytime peak value of $ \approx 50 $ TECU was attained at 1400 UT while for the observed TEC profile, a daytime peak of $ \approx 60 $ TECU was reached at that same time. Between midnight and 0700 UT on September 26, $TEC_{obs} $ values and $TEC_{qav} $ values were observed to be very close. Similar observation was made between 1800 UT and 1900 UT. Except for those hours, profound TEC enhancements were noticed at remaining hours of September 26; the TEC profile showed that $TEC_{obs} $ values were varied significantly larger than the $TEC_{qav} $ between 0700 UT and 1600 UT before the main phase of the storm commenced and initially between 1600 UT and 1800 UT and later between 1900 UT and mid-night as the storm main phase lasted. The maximum variation was ob-served at 2000 UT during the main phase of the storm event. 

As reported by \cite{Dashora2005}, scintillation values must be greater than or equal to 0.3 for a significant level of scintillation phenomena. In the case of this storm, during the main phase hours (i.e., be-tween 1800 UT to 0000 UT), the scintillation index recorded minimum and maximum values of 0.05 and 0.39 respectively. Between the hours of 1800 UT and 2300 UT on September 26, the maximum scintillation index value was 0.27 at 2300 UT after which the value of 0.39 was recorded at 0000 UT on September 27. Clearly, a less significant scintillation phenomenon had occurred during the September 26 -- 27, 2011 storm event when compared with the August 5 -- 6, 2011 geomagnetic storm event. Implicatively, the occurrence of a geomagnetic storm event does not necessarily suggest an increase in the level of scintillation at a low–latitude region. A similar observation was reported by \cite{Ariyibi2013}.

\subsection{Interplanetary Effects on Main Phase of the Storm Events}
\label{subsec3}
According to \cite{Gonzalez1994}, the main phase is the principal feature of the Dst representation of a geomagnetic storm. The depression of the magnetic field during the main phase is explained as the effect of the ring current in the magnetosphere. The ring current is carried primarily by energetic (20 -- 200 keV) ions \cite{Kamide2001}. Deductively, the nature of the main phase of a geomagnetic storm provides an explanation of the process of the ring current energization responsible for the development of the storm. The two storms under consideration are both characterized by a maximum depression of the Dst Index exceeding 100 nT, this is indicative of a strong ring current intensification. Large-scale IEF driven by a period of prolonged southward IMF-Bz may be responsible for the ring current build-up. 

The development of the main phase of the two geomagnetic storm events was clearly associated with the southward turning of the inter-planetary magnetic field. For the periods under consideration, the depressions of the Dsts were corresponded by southwardly directed IMF-Bz. This demonstrates that a strong southward interplanetary magnetic field is sufficient for the intensification of the ring current. In both cases, an abrupt increase in solar wind speed confirmed the presence of a shock in the interplanetary medium \cite{Kelley2009} as the main phase of the geomagnetic storm events commenced. The Kp values of $ \geq $ 7 for both storms also confirmed the intensity of the geomagnetic activity. 

\cite{Huang2008} reported that penetration electric field lasts for eight to ten hours without obvious attenuation during the main phase of a magnetic storm when the IMF remains southward. IMF-Bz for the August 5 -- 6, 2011 storm event turned southward at 2000 UT and remained in that direction until 0400 UT, a period of eight hours while IMF-Bz for the September 26 -- 27, 2011 storm event turned southward at 1600 UT and remained in that direction until 0000 UT, a duration of eight hours. 

This brings to focus the effects of PPEF, originating due to under and over shielding conditions, on equatorial and low latitude ionosphere \cite{Dashora2009a}. The peak values recorded by IMF-Bz for both storms are indicative of strong penetration electric fields. The rapid decrease in IMF-Bz during the main phase of the storm events causes a sudden decrease in high latitude electric potential, resulting in the generation of sudden region-1 currents. The region-2 current, the lower latitude current system which shields charges generated by the ring current cannot respond at the same rate. The high latitude electric potential which is normally shielded by the region-2 currents was able to reach much lower latitudes \cite{Peymirat2000}. This condition is referred to as undershielding, which results in prompt penetration of interplanetary electric fields to low latitudes and equator. Implicatively, the shielding process of the region-2 field-aligned current was not effective during the main phase of both geomagnetic storms.

Furthermore, \cite{Kamide2001} noted that during the largest magnetic storms, the main phase often goes through two steps. That is, well be-fore the Dst decrease has fully recovered to its pre-storm level, a succeeding decrease tends to occur. For the periods under consideration, the occurrence of a two-step main phase was observed in the August 5 -- 6, 2011 storm event, but a triple-step development was observed in the September 26 -- 27, 2011 storm event. This would mean that the geomagnetic storms consisted merely of two and three medium-size storms respectively. By implication, what is perhaps needed in generating an in-tense magnetic storm are two (or three) separate regions of southward IMF in an interplanetary structure. One of the most important factors in generating intense magnetic storms is perhaps the slow decay rate of the ring current. 

\subsection{Storm Main Phase Effect on Ionospheric TEC and $ S_4 $}
\label{subsec4}
The diurnal pattern of the total electron content for the periods under consideration exhibited a steady increase from about sunrise to an after-noon maximum and then falls to attain a minimum during the sunset and post-sunset hours, this pattern is typical of low-latitude ionospheric stations \cite{Bagiya2009}. The TEC curves show appreciable day-to-day variations of TEC, particularly during the mid-day which is a serious concern in forecasting and navigation \cite{Bagiya2009,Rao2009}. This variation may be attributed to the changes in solar activity. Large variations in TEC are observed in daytime while night-time variations were observed to be almost constant. 

Compared with TEC values at the other two phases of the geomagnetic storm, TEC values during the main phase fell by about 80\%. TEC values for the post-sunset hours of August 6 were about double the TEC values for post-sunset hours during the main phase hours on August 5, also, TEC values during pre-sunrise hours for August 5 were about dou-ble the TEC value for the pre-sunrise hours during the main phase hours on August 6. While TEC depletion was observed in the case of the August 5 -- 6, 2011 storm, TEC enhancement was observed in the case of the September 26 – 27, 2011.

Generally, an increase in storm time total electron content from the average quiet values was observed for both storms. According to \cite{Kikuchi1996}, when z-component of the IMF-Bz suddenly turns southwards from a steady northward configuration, a dawn to dusk convection electric field is produced at high latitudes. With sudden southward turning of IMF-Bz observed during both storms, increase in TEC value can be attributed to this high latitude convection electric field penetrating to the equatorial and low latitudes. The effect of the main phase of the geomagnetic storm events on TEC was significant such that large deviations from quiet-time average values were noticeable. The variation was observed most at the decay region of TEC for both periods considered.

$ S_4 $ recorded values ranging between minimum and maximum values of 0.0 and about 0.57 respectively during the main phase of the August 5 -- 6, 2011 storm event while the $ S_4 $ recorded minimum and maximum values of 0.05 and 0.39 respectively during the main phase of the September 26 -- 27, 2011 storm event. A remarkable scintillation suppression was observed in August which is most likely associated with the effect due to prompt penetration of an eastward electric field into the equatorial ionosphere (Priyadarshi and Singh, 2011). It is however clear from the September storm event that geomagnetic disturbance may not trigger the scintillation occurrence in low latitude regions.

Implicatively, as seen from the Dsts, the ring current was more intensified and was prolonged for more hours in the August 5 – 6, 2011 storm event (two minimum Dsts of –88 nT and –107 nT for four hours) than in the September 26 – 27, 2011 storm event (three minimum Dsts of –93 nT, –100 nT and –101 nT for three hours), this may explain why there was more pronounced scintillation effect in August than in September.

\section{Conclusion}
\label{sec4}
The condition of the equatorial ionosphere over Ile-Ife during the main phase of two intense geomagnetic storm events of August 5 -- 6, 2011 and September 26 -- 27, 2011 was investigated in this study using a combination of GPS data obtained at the Ile-Ife station and interplanetary and geomagnetic data. Noticeably, the main phase of both storms occurred at night-time with one of them having its main phase cutting across midnight. 

The intensity of a geomagnetic storm event correlates well with a strong southward direction of the IMF-Bz and the Vsw. The main phase of intense geomagnetic storm events is characterized by two or three step development which consists of two or three moderate-sized storms. One of the most important factors responsible for this observation is perhaps the slow decay rate of the ring current. The main phase of the geomagnetic storm event caused TEC depletion on August 5 -- 6, 2011 while TEC enhancements were observed on September 26 -- 27, 2011.

Enhancement in TEC has been attributed to local low latitude response to dawn-to-dusk penetration electric fields while depletions have been attributed to, either the termination of the penetration eastward electric field or to the penetration of westward (dusk-to-dawn) electric field to low latitudes \cite{Dashora2009a}. When the main phase of the storm event on August 5 -- 6, 2011 commenced around 2100 UT, scintillation was quite pronounced, recording a maximum value of 0.57 and corresponding to rapid fluctuations in TEC, as the main phase progressed, significant scintillation suppression followed for about four hours.
\vspace{5mm}

\textbf{Acknowledgements}. \textit{We express our gratitude to Dr. C. Carrano, Dr. P. Doherty and Dr. S. Gopi Krishna (Institute of Scientific Research, Boston College, Chestnut Hill, USA) who provided the GPS-SCINDA Sensor System and the GPSTEC Analysis Software that was used for this research. We also acknowledge the World Geomagnetic Data Center, Kyoto, Japan and Space Physics Data Facility of National Aeronautics and Space Administration, USA for the datasets acquired for this study.}

\textbf{Funding}. \textit{This research did not receive any specific grant from funding agencies in the public, commercial, or
not-for-profit sectors.}

\end{document}